\documentclass[reprint,aps,superscriptaddress,prl]{revtex4-1}

\usepackage{graphicx}
\usepackage{amssymb}
\usepackage{amsmath}
\usepackage{color}
\usepackage{float}
\usepackage{multirow}

\newcommand{\beq}{\begin {equation}}
\newcommand{\eeq}{\end   {equation}}
\newcommand{\bea}{\begin {eqnarray}}
\newcommand{\eea}{\end   {eqnarray}}
\newcommand{\baa}{\begin {array}   }
\newcommand{\eaa}{\end   {array}   }
\newcommand{\bit}{\begin {itemize} }
\newcommand{\eit}{\end   {itemize} }
\newcommand{\be }{\begin {equation}}
\newcommand{\ee }{\end   {equation}}
\newcommand{\nn }{\nonumber        }
\newcommand{\tabincell}[2]{\begin{tabular}{@{}#1@{}}#2\end{tabular}}

\begin{document}

\title{Probing $HZ\gamma$ and $H\gamma\gamma$ anomalous couplings in the process of $e^+e^- \to H\gamma$}

\author{Qing-Hong Cao}
\email{qinghongcao@pku.edu.cn}
\affiliation{Department of Physics and State Key Laboratory of Nuclear Physics
and Technology,
\\Peking University, Beijing 100871, China}
\affiliation{Collaborative Innovation Center of Quantum Matter, Beijing, China}
\affiliation{Center for High Energy Physics, Peking University, Beijing 100871, China}

\author{Hao-Ran Wang}
\email{haorwang@pku.edu.cn}
\affiliation{Department of Physics and State Key Laboratory of Nuclear Physics
and Technology,
\\Peking University, Beijing 100871, China}

\author{Ya Zhang}
\email{yazhang@pku.edu.cn}
\affiliation{Department of Physics and State Key Laboratory of Nuclear Physics
and Technology,
\\Peking University, Beijing 100871, China}

\begin{abstract}
Rare decay of the Higgs boson is a powerful tool to probe new physics beyond the standard model. The rare decays occur through quantum loops in which the standard model and new physics contributions might cancel each other out. We consider a faked-no-new-physics scenario that the new physics contributions are about minus two times the standard model contribution such that partial widths of Higgs boson rare decays are the same as the standard model predictions.  We propose to measure the $HZ\gamma$ and $H\gamma\gamma$ anomalous couplings in the process of $e^+ e^- \to H\gamma$ that is strongly correlated to Higgs boson rare decays. We show that the faked-no-new-physics scenario can be fully probed at a high energy electron-positron collider.
\end{abstract}

\maketitle

One of the major tasks of particle physics is to precisely measure the Higgs boson property.
Rare decays of the Higgs boson, $H\to Z\gamma$ and $H\to \gamma\gamma$, offers a rich potential for new physics  (NP) searches. Observing a deviation from the standard model (SM) prediction would shed light on NP models. 
The $H\gamma\gamma$ and $HZ\gamma$ anomalous couplings are sensitive to different kind of NP and therefore are independent in principle. Measuring the $HZ\gamma$ and $H\gamma\gamma$ couplings accurately is useful to test and discriminate NP models. For example, the $HZ\gamma$ coupling could be sizably modified in certain composite Higgs model while keeping the $H\gamma\gamma$ coupling unchanged~\cite{Azatov:2013ura}.  On the contrary, the two anomalous couplings are highly correlated in NMSSM or MSSM-like~\cite{Gainer:2011aa,Carena:2012xa, Cao:2013ur, Arhrib:2014pva,Belanger:2014roa,Hu:2014eia}. A large derivation in the $HZ\gamma$  coupling from the SM expectation, which is not correlated with a similar deviation in the $H\gamma\gamma$ coupling, would impose strong constraints on NP models.

Our knowledge of the $HZ\gamma$ and $H\gamma\gamma$ couplings are obtained from branching ratios or partial decay widths of Higgs boson rare decays. It is commonly believed that the NP effect is small if the branching ratio measurement is well consistent with the SM prediction. The rare decays of Higgs bosons occur through quantum loops in which the SM and NP contributions might cancel each other out. In this Letter, we focus on a {\it faked-no-new-physics} scenario that the NP contribution is about minus two times the SM contribution.  Such a scenario yields similar partial widths of Higgs boson rare decays as the SM predictions and cannot be tested in Higgs boson decays at the Large Hadron collider (LHC). We show that the $H\gamma$ production in the electron-positron collision can verify or exclude the hidden NP effects.  Measurements of the $H\gamma$ production cross section and the partial decay widths $\Gamma(H\to Z\gamma/\gamma\gamma)$  are especially valuable because their correlation can be related through the anomalous couplings to the structure of NP models. 

We begin with a general assumption that effects beyond the SM are described by a set of higher dimensional operators made out of the SM fields only. At the level of effective Lagrangian, effective operators after spontaneously symmetry breaking contribute to the $HZ\gamma$ and $H\gamma\gamma$ anomalous couplings as follows: 
 \beq
 \mathcal{L}=\frac{v}{\Lambda^{2}}\left(\mathcal{F}_{Z\gamma} HZ_{\mu\nu}A^{\mu\nu} + \mathcal{F}_{\gamma\gamma}H A_{\mu\nu}A^{\mu\nu}\right)\nn
\eeq
where $v=246~{\rm GeV}$ is the vacuum expectation value while $\Lambda$ the NP scale. Throughout this work we choose $\Lambda=2~{\rm TeV}$.   The coefficients $\mathcal{F}_i$ are real to respect the CP parity.  
One can probe the anomalous couplings at the LHC from the branching ratio of $H \to \gamma\gamma$ and $H\to Z\gamma$ rare decays. The partial decay widths are  
\bea
\Gamma(H\rightarrow Z\gamma) &=& \frac{m_{H}^{3}}{8\pi v^{2}} \left(1-\frac{m_{Z}^{2}}{m_{H}^{2}}\right)^3
\biggl|\mathcal{F}_{Z\gamma}^{\rm SM}+\frac{v^2}{\Lambda^2} \mathcal{F}_{Z\gamma}\biggr|^2 \nn\\
\Gamma(H\rightarrow \gamma\gamma) &=& \frac{m_{H}^{3}}{16\pi v^2}
\biggl|\mathcal{F}_{\gamma\gamma}^{\rm SM}+\frac{v^2}{\Lambda^2} \mathcal{F}_{\gamma\gamma}\biggr|^2,
\eea
where $\mathcal{F}_{Z\gamma}^{\rm SM}$ and $\mathcal{F}_{\gamma\gamma}$, induced by the $W$-boson and top-quark loops in the SM, are given by~\cite{Azatov:2013ura, Low:2012rj}
\bea
\mathcal{F}_{Z\gamma}^{\rm SM} &=& \frac{\alpha}{4\pi s_{W}c_W} \biggl(\hat{v}_f A_{1/2}^H(\tau_{t},\lambda_{t})+ c_{W}A_{1}^H(\tau_{W},\lambda_{W})\biggr)， \nn\\
\mathcal{F}_{\gamma\gamma}^{\rm SM} &=& \frac{\alpha}{4\pi } \biggl( 3Q_{t}^{2}A^{H}_{1/2}(\tau_{t}^{-1})+A^{H}_{1}(\tau_{W}^{-1})\biggr),
\eea
where the functions $A_{1/2}^H$ and $A_1^H$ are given in Ref.~\cite{Djouadi:2005gi} with $\tau_i=4m_i^2/m_H^2$ and $\lambda_i=4m_i^2/m_Z^2$. $\hat{v}_f = N_c Q_{t}(2T_{3}^{t}-4Q_{t}s^{2}_{W})$ with $Q_t$ being the top-quark electric charge in units of $|e|$ and $T_3^t=1/2$. 

Note that the $W$-bosons and top-quarks inside the loop of the Higgs boson rare decay cannot be on-shell for $m_H=125~{\rm GeV}$. As a result, both $\mathcal{F}_{Z\gamma}^{\rm SM}$ and $\mathcal{F}_{\gamma\gamma}^{\rm SM}$ are real and their values are 
\beq
\mathcal{F}_{Z\gamma}^{\rm SM}\sim 0.007,~~\mathcal{F}_{\gamma\gamma}^{\rm SM}\simeq -0.004~.
\eeq
Here the bottom-quark loop contribution is ignored. The ATLAS and CMS measurements impose bounds on the signal strength relative to the SM prediction as follows~\cite{Aad:2014eha,Khachatryan:2014ira,ATLASHzg2014,CMSHzg2013}:
\begin{align}
&R_{Z\gamma}\equiv \frac{\Gamma(H\rightarrow Z\gamma)}{\Gamma_{\rm SM}(H\rightarrow Z\gamma)}\leq 9.5~\nn\\
0.91\leq &R_{\gamma\gamma}\equiv \frac{\Gamma(H\rightarrow \gamma\gamma)}{\Gamma_{\rm SM}(H\rightarrow \gamma\gamma)}\leq 1.4
\end{align}
at the 95\% confidence level. It yields a bound on $\mathcal{F}_{Z\gamma}$ as $-2.02\leq \mathcal{F}_{Z\gamma}\leq 1.03$ and two bounds on $\mathcal{F}_{\gamma\gamma}$ as $-0.051 \leq \mathcal{F}_{\gamma\gamma}\leq 0.013$ and $0.55 \leq \mathcal{F}_{\gamma\gamma}\leq 0.62$. 

There are two-fold solutions of $\mathcal{F}_{Z\gamma,\gamma\gamma}$ for each fixed value of the partial decay widths. The ambiguity cannot be resolved in the branching-ratio measurements of Higgs boson rare decays. 
In particular, $R_i\sim 1$ for $\mathcal{F}_i \sim 0$ (no-new-physics) or $v^2/\Lambda^2 \mathcal{F}_i \sim -2\mathcal{F}_{i}^{\rm SM}$ (fake-no-new-physics scenario).
We propose to determine both magnitude and sign of the $HZ\gamma$ and $H\gamma\gamma$ couplings in the process of $e^+ e^- \to Z^*/\gamma^* \to H \gamma$ at future electron-position colliders. Different from Higgs boson rare decays, the SM amplitudes of the $H\gamma$ production develop imaginary parts to enhance the production rate~\cite{Gounaris:2015tna}. That helps to resolve the two-fold solutions.

\begin{figure}
\includegraphics[scale=0.3,clip]{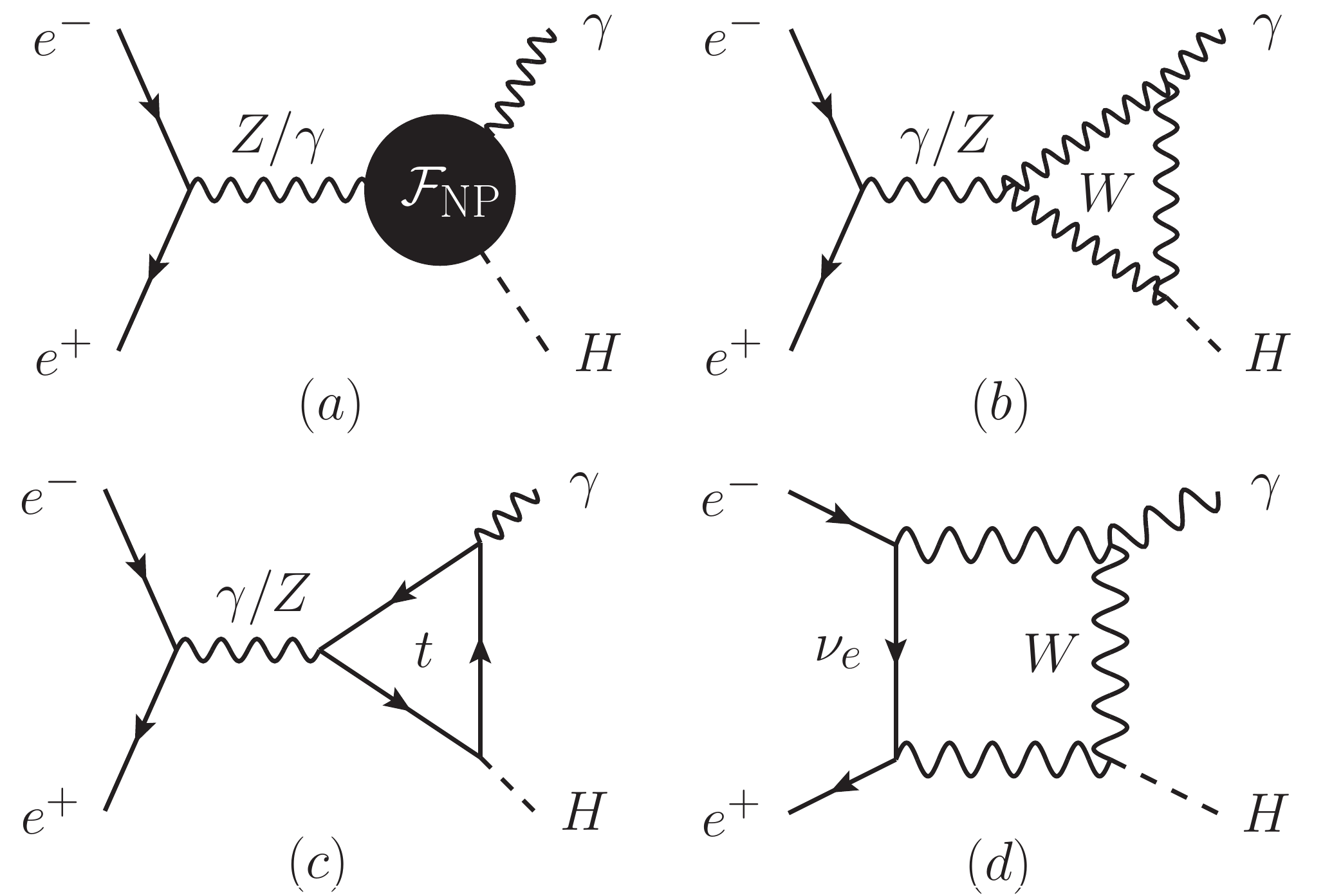}
\caption{\it Representative Feynman diagrams of $e^{+}e^{-}\rightarrow H \gamma$: the anomalous couplings  (a) and SM diagrams (b-d).}
\label{fig:feyn}
\end{figure}

The scattering process of $e^+e^- \to H\gamma$ is absent at the tree-level in the SM when ignoring the electron mass, but it can be generated through the electroweak corrections at the loop-level~\cite{Barroso:1985et, Abbasabadi:1995rc, Djouadi:1996ws}. The effects of the $HZ\gamma$ and $H\gamma\gamma$ anomalous couplings might be comparable to those SM loop effects. One has to consider the SM loop contributions as well in the discussion of the NP effects. Figure~\ref{fig:feyn} displays the representative Feynman diagrams of both NP and SM contributions.

We calculate the SM loop corrections in FormCalc~\cite{Hahn:1998yk} and LoopTools~\cite{vanOldenborgh:1990yc}. Our analytical and numerical results are consistent with those in Refs.~\cite{Djouadi:1996ws}. We then incorporate the $HZ\gamma$ and $H\gamma\gamma$ anomalous couplings into our calculation. 
In order to quantify the NP effects, we separate the total cross section of the $H \gamma$ production ($\sigma_{\rm t}$) into the following three pieces:
\bea
\label{smnp}
  \sigma_{\rm t}&=&\sigma_{\rm SM}+\left[ \sigma_{\rm IN}^{(1)} \mathcal{F}_{Z\gamma} +\sigma_{\rm IN}^{(2)} \mathcal{F}_{\gamma\gamma} \right] \left(\tfrac{\rm 2TeV}{\Lambda}\right)^2 \nn\\
  &+& \left[\sigma_{\rm NP}^{(1)} \mathcal{F}_{Z\gamma}^2+ \sigma_{\rm NP}^{(2)} \mathcal{F}_{\gamma\gamma}^2 + \sigma_{\rm NP}^{(3)} \mathcal{F}_{Z\gamma}\mathcal{F}_{\gamma\gamma} \right]\left(\tfrac{\rm 2TeV}{\Lambda}\right)^4 , 
\label{eq:xsection}
\eea
where $\sigma_{\rm SM}$ is the SM cross section, $\sigma_{\rm IN}^{(1,2)}$ is the interference effects between the SM and NP contributions and $\sigma_{\rm NP}^{(1,2,3)}$ is the NP contribution.  For illustration we list the total cross section (in the unit of femtobarn) for $\sqrt{s}=350~{\rm GeV},~500~{\rm GeV}$ and $1000~{\rm GeV}$ as follows:
\begin{widetext}
\begin{eqnarray}
350~{\rm GeV}&:&\sigma_{\rm t}= 0.0341 +\left[0.2524 \mathcal{F}_{Z\gamma}+0.0105 \mathcal{F}_{\gamma\gamma}\right]\left(\tfrac{\rm 2 TeV}{\Lambda}\right)^2 + \left[ 0.5212  \mathcal{F}_{Z\gamma}^2 +  1.2392\mathcal{F}_{\gamma\gamma}^2 +  0.1750\mathcal{F}_{Z\gamma}\mathcal{F}_{\gamma\gamma} \right]\left(\tfrac{\rm 2TeV}{\Lambda}\right)^4, \nn\\
500~{\rm GeV}&:&\sigma_{\rm t}= 0.0524+\left[0.2865 \mathcal{F}_{Z\gamma}+0.3613 \mathcal{F}_{\gamma\gamma}\right]\left(\tfrac{\rm 2 TeV}{\Lambda}\right)^2 + \left[ 0.6012  \mathcal{F}_{Z\gamma}^2 +  1.5375\mathcal{F}_{\gamma\gamma}^2 + 0.2093\mathcal{F}_{Z\gamma}\mathcal{F}_{\gamma\gamma} \right]\left(\tfrac{\rm 2TeV}{\Lambda}\right)^4,\nn\\
1000~{\rm GeV}&:&\sigma_{\rm t}= 0.0214+\left[0.1703 \mathcal{F}_{Z\gamma}+0.2808\mathcal{F}_{\gamma\gamma}\right]\left(\tfrac{\rm 2 TeV}{\Lambda}\right)^2 + \left[0.6614\mathcal{F}_{Z\gamma}^2 +  1.7799\mathcal{F}_{\gamma\gamma}^2 + 0.2362\mathcal{F}_{Z\gamma}\mathcal{F}_{\gamma\gamma} \right]\left(\tfrac{\rm 2TeV}{\Lambda}\right)^4.\nn\\
\label{eq:xsection2}
\end{eqnarray}
\end{widetext}

\begin{figure}
\includegraphics[scale=0.24]{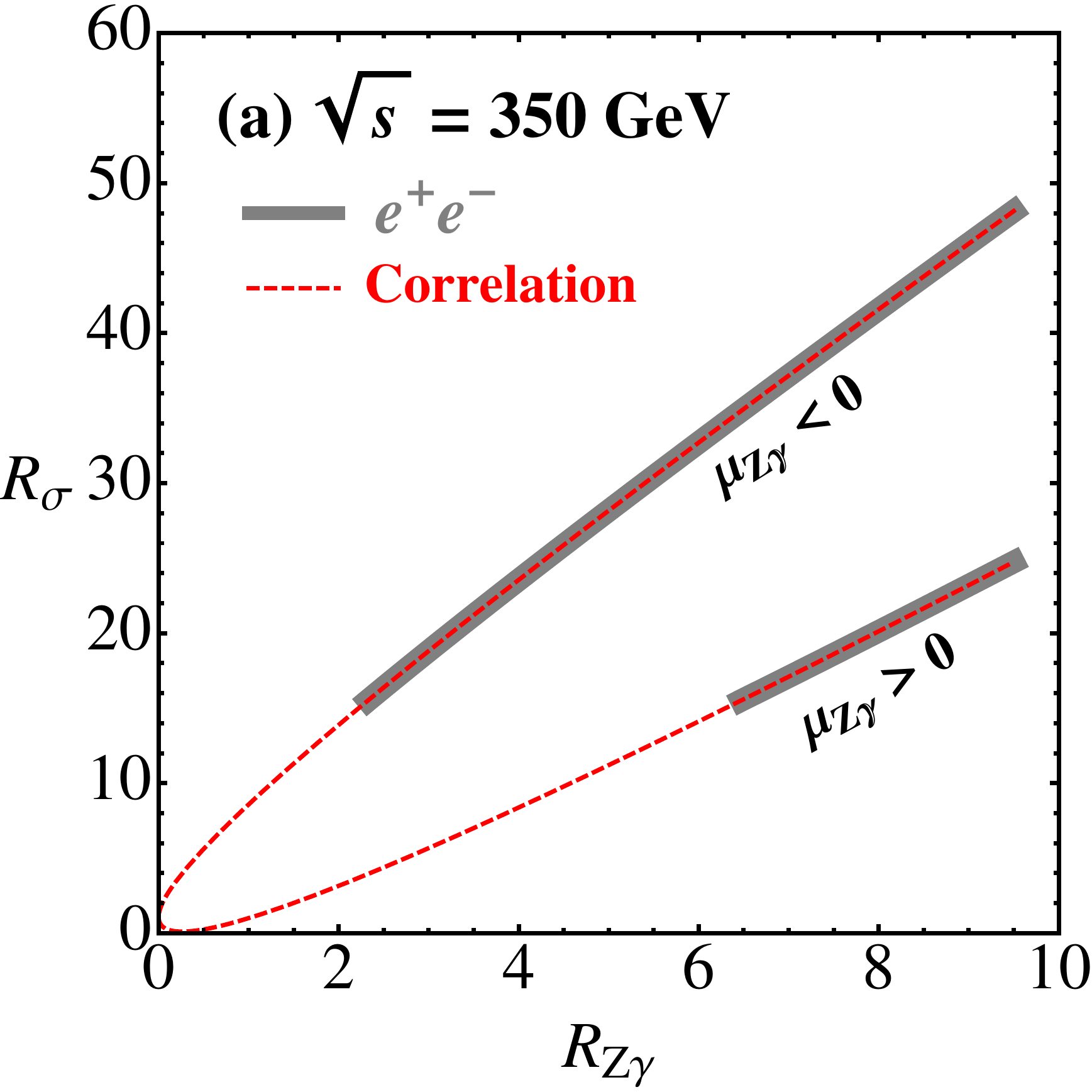}
\includegraphics[scale=0.24]{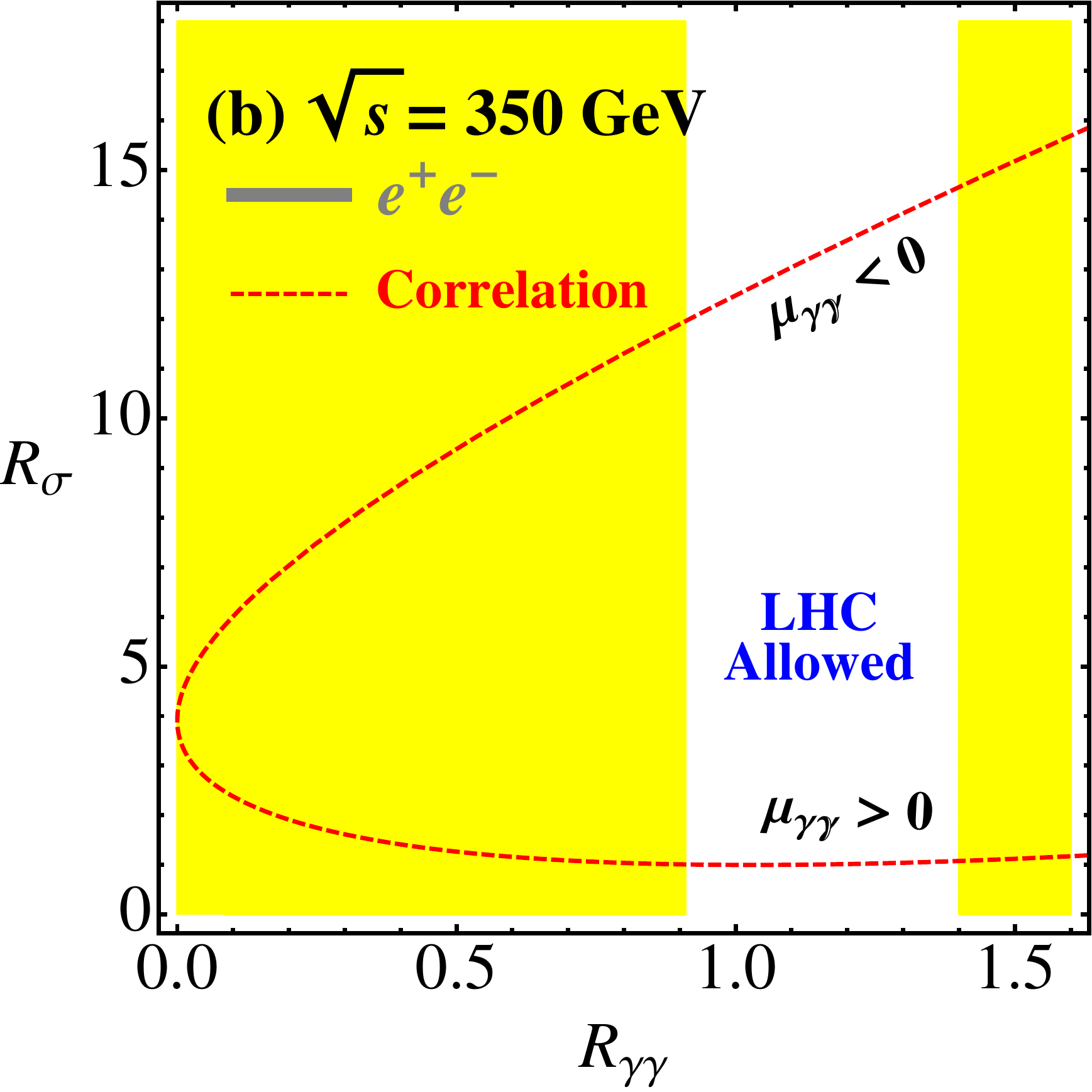}\\
\includegraphics[scale=0.24]{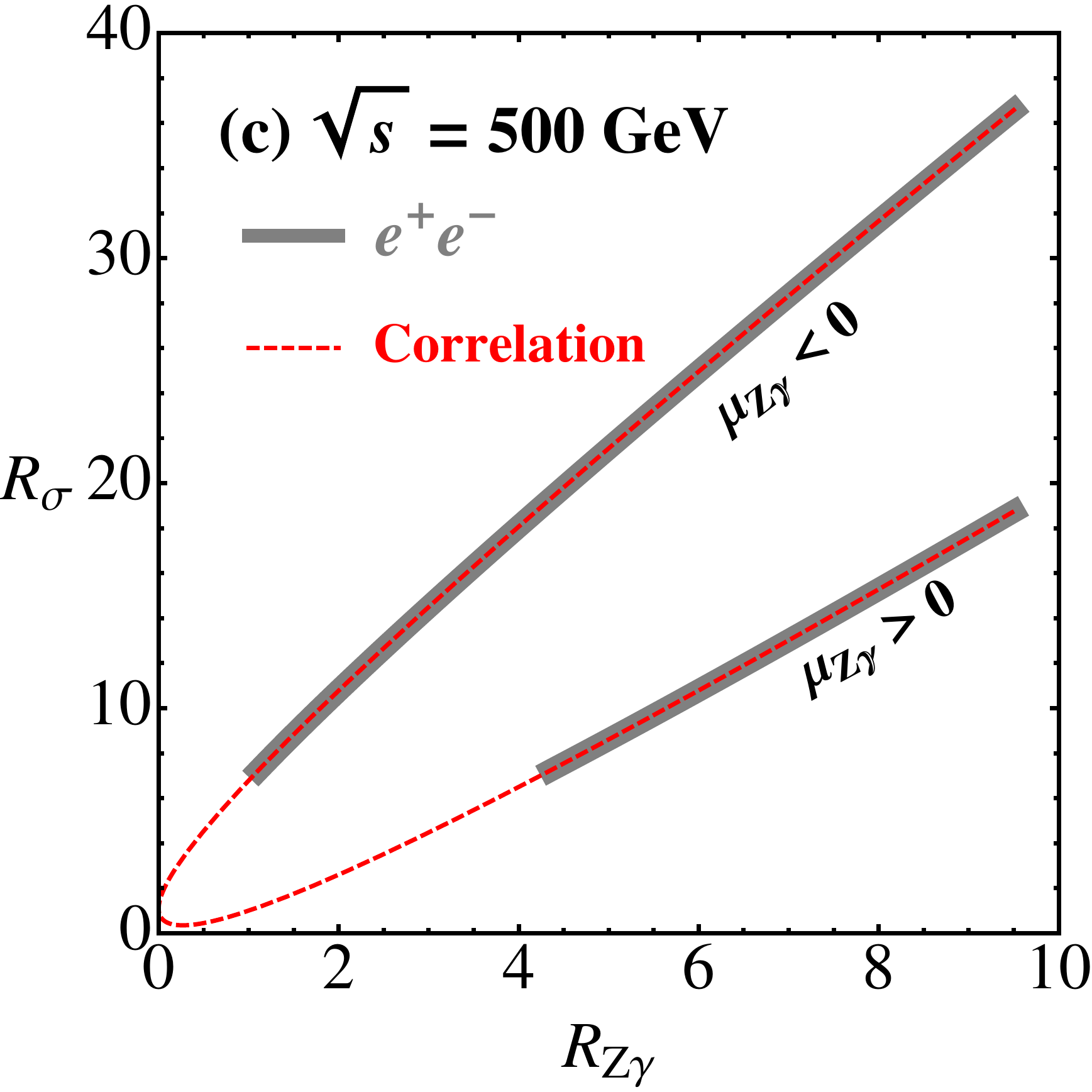}
\includegraphics[scale=0.24]{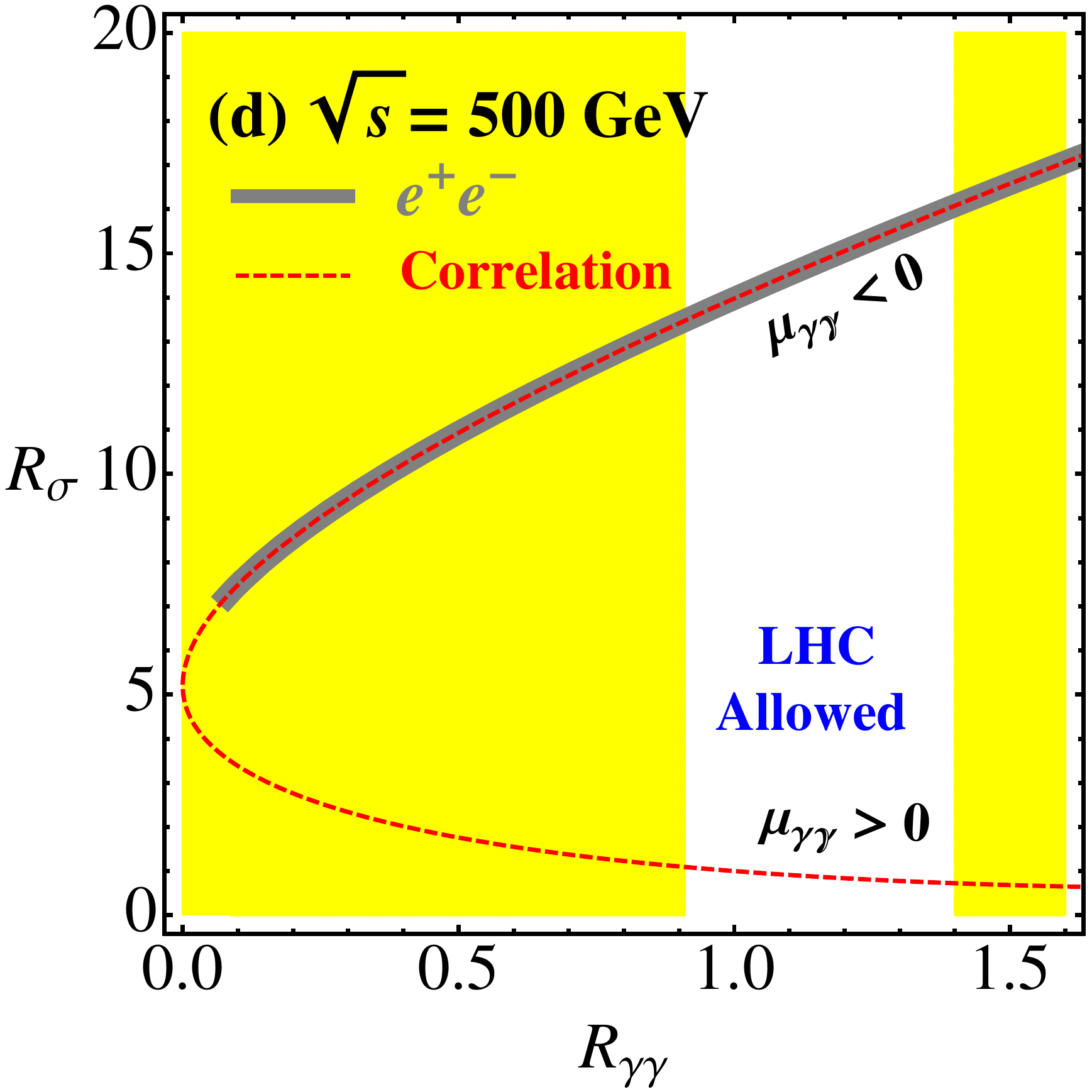}\\
\includegraphics[scale=0.24]{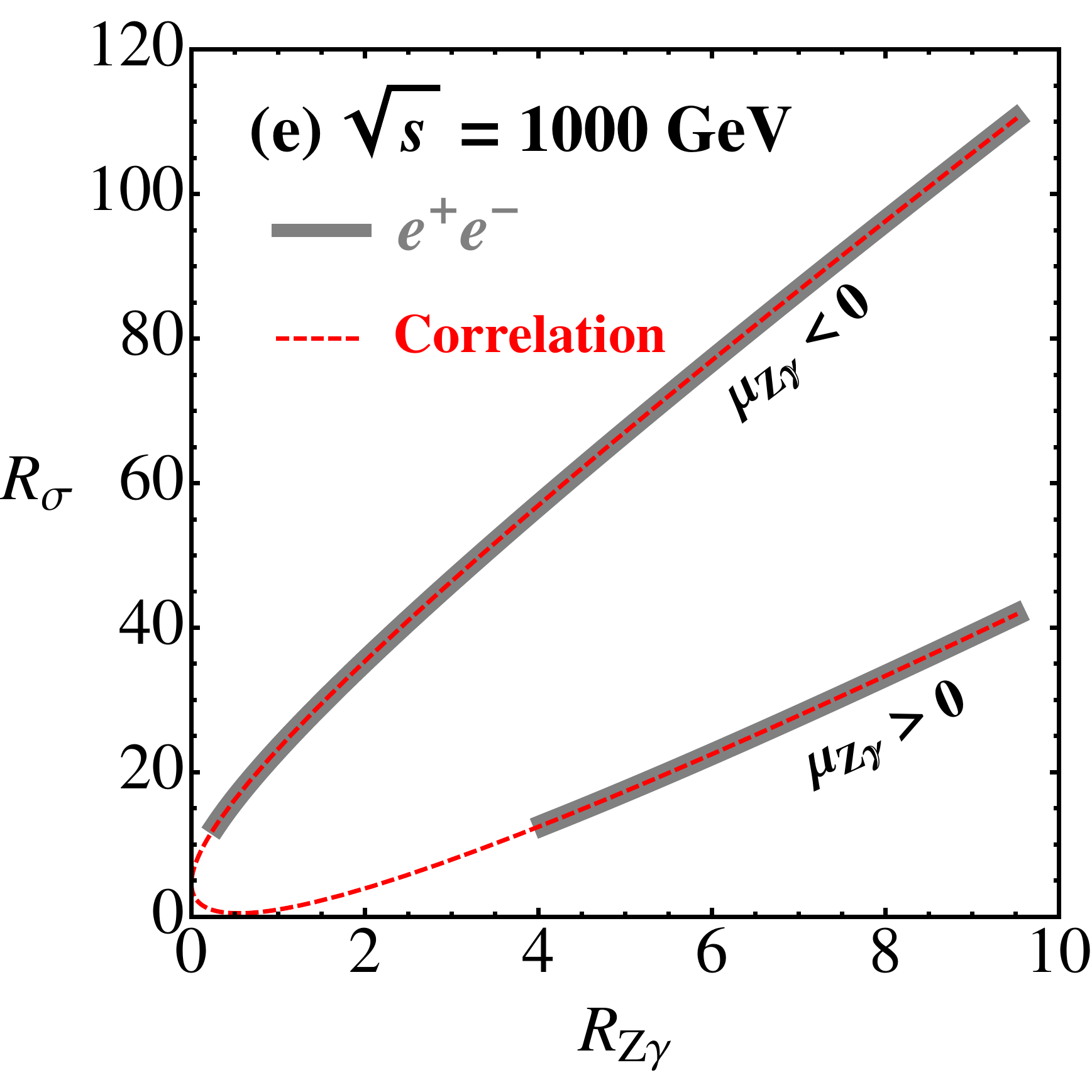}
\includegraphics[scale=0.24]{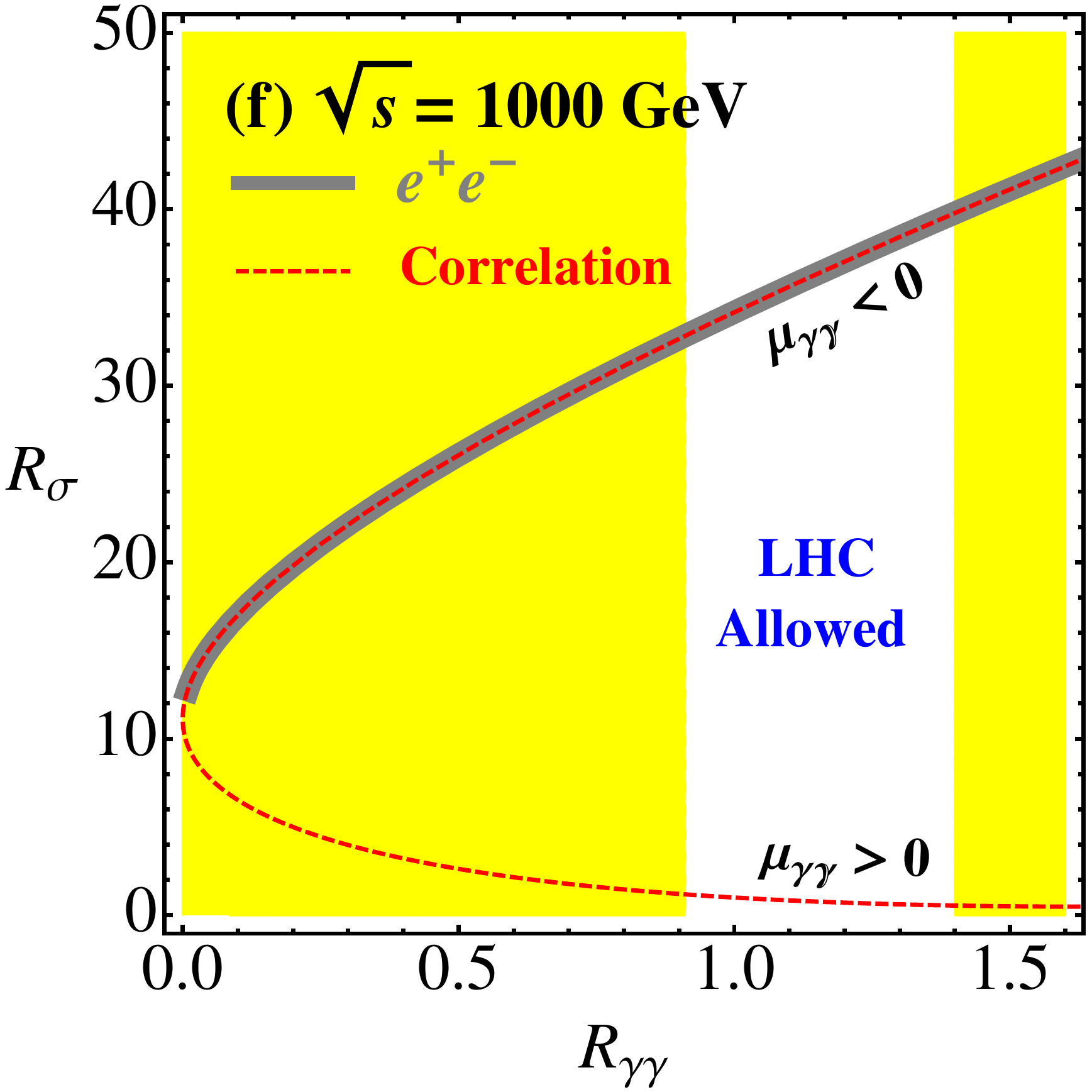}\\
\caption{\it Correlations between  $R_\sigma$ and $R_{Z\gamma/\gamma\gamma}$ (red-dashed line) and discovery region at the $e^+e^-$ colliders (bold-gray curve). The yellow shadow regions are excluded by recent LHC data. One anomalous coupling is considered at a time. }
\label{fig:corr2}
\end{figure}

The anomalous couplings generate strong correlations among the $H\gamma$ production and Higgs boson rare decays. 
Figure~\ref{fig:corr2} displays the strong correlation of $R_\sigma$ and $R_{Z\gamma/\gamma\gamma}$ (red-dashed curves) where $R_{\sigma}$, $R_{Z\gamma/\gamma\gamma}$ and the relative sign $\mu_{Z\gamma/\gamma\gamma}$ are defined as follows:
\begin{align}
&R_{\sigma} \equiv \frac{\sigma_{\rm t}(e^+e^- \to H\gamma)}{\sigma_{\rm SM}(e^+e^- \to H\gamma)},&&\nn\\
&R_{Z\gamma} \equiv \frac{\Gamma(H\to Z\gamma)}{\Gamma_{\rm SM}(H\to Z\gamma)},
&&\mu_{Z\gamma}={\rm sign}\left(\frac{\mathcal{F}_{Z\gamma}}{\mathcal{F}^{\rm SM}_{Z\gamma}}\right),\nn\\
&R_{\gamma\gamma} \equiv \frac{\Gamma(H\to \gamma\gamma)}{\Gamma_{\rm SM}(H\to \gamma\gamma)},& &\mu_{\gamma\gamma}={\rm sign}\left(\frac{\mathcal{F}_{\gamma\gamma}}{\mathcal{F}^{\rm SM}_{\gamma\gamma}}\right).
\end{align}
There are two values of $R_\sigma$ for each fixed $R_{Z\gamma/\gamma\gamma}$; the larger $R_\sigma$ corresponds to $\mu_{Z\gamma/\gamma\gamma}<0$ while the smaller to $\mu_{Z\gamma /\gamma\gamma}>0$.

Now we discuss how to detect the Higgs anomalous coupling at the $e^+e^-$ collider. We focus on the $b\bar{b}$ mode of Higgs boson decay. The collider signature of interest to us is one hard photon and two $b$-jets.  We choose $\mathcal{F}_{Z\gamma}=1$ to model the kinetics of the anomalous couplings. The cut efficiencies obtained also apply to other values of $\mathcal{F}_{Z\gamma / \gamma\gamma}$. The dominant SM background, $ e^+  e^- \to \gamma\gamma^*/ \gamma Z \to \gamma  b \bar{b}$, is generated in MadGraph~\cite{Alwall:2014hca}. 
At the analysis level, all signal and background events are required to pass the  {\it selection cuts}:
\beq
p_T^{\gamma, b, \bar{b}} >  25~{\rm GeV}, ~|\eta^{\gamma, b, \bar{b}} | \leq 3.5, ~ \Delta R_{b\bar{b}, b\gamma, \bar{b}\gamma} \geq 0.7, 
\eeq
where $p_T^i$ and $\eta^i$ denotes the transverse momentum and pseudo-rapidity of the particle $i$, respectively. The separation $\Delta R$ in the azimuthal angle ($\phi$)-pseudo-rapidity ($\eta$) plane between the objects $k$ and $l$ is
$\Delta R_{kl}\equiv \sqrt{(\eta_k-\eta_l)^2 + (\phi_k - \phi_l)^2}$.
For simplicity we ignore the effects due to the finite resolution of the detector and assume a perfect $b$-tagging efficiency.

Table~\ref{tab:cut} shows the rates of the signals ($\mathcal{S}$) and backgrounds ($\mathcal{B}$)  before and after cuts, with $\mathcal{F}_{Z\gamma}=1$, for four values of the c.m. energy. We assume an integrated luminosity of $1~{\rm ab}^{-1}$.  
The numbers of the signal and background events after imposing the above selection cuts are summarized in the second  and fifth rows of Table~\ref{tab:cut}. The signal consisting of both the SM and NP contributions is shown in the fourth to sixth rows.  Obviously, the backgrounds are larger than the signals by three or four order of magnitudes. One has to impose other cuts to extract the small signal out of the huge background.

\begin{table}
\caption{\it The number of events of the signal ($\mathcal{S}$) and the background ($\mathcal{B}$) for various c.m. energies ($\sqrt{s}$) with an integrated luminosity of $1~{\rm ab}^{-1}$.  ${\rm Br}(H\to b\bar{b})=74.8\%$ is included. For illustration we choose  $\mathcal{F}_{Z\gamma}=1$ and $\mathcal{F}_{\gamma\gamma}=0$. }
\label{tab:cut}
  \begin{tabular}{c|c|c|c|c|c}
\hline
 \multicolumn{2}{c|}{$\sqrt{s}$ (GeV)}& 250 &350 & 500&1000\tabularnewline
\hline
\hline
\multirow{2}{*}{$\mathcal{B}$}&{\it selection  cuts} ($\times 10^{5}$) & 7.169 & 4.229 & 2.450 &0.708 \tabularnewline
\cline{2-6}
& $E_r$, $\Delta M$ {\it cut} & 7640 & 3993 & 2104&475\tabularnewline
\hline
\hline
\multirow{3}{*}{\tabincell{c}{~\\$\mathcal{S}$\\~\\~ }}
&{\it Inclusive rate} & 594 & 605 & 703 & 638\tabularnewline
\cline{2-6}
&{\it selection cuts} & 451 & 482 & 569 & 341\tabularnewline
\cline{2-6}
& $E_r$, $\Delta M$  {\it cut} & 451 & 482 & 569 & 341\tabularnewline
\cline{2-6}
\hline
\multicolumn{2}{c|}{$\mathcal{S}/\sqrt{\mathcal{B}}$}& 5.2  & 7.6 & 12.4 & 15.6\tabularnewline
\hline
\end{tabular}
\end{table}

The photon exhibits a recoil energy to balance the Higgs boson production,  $E_\gamma = (s-m_H^2)/2\sqrt{s}$, which could be used to trigger the signal events. The two $b$-jets in the signal originate from the Higgs boson while those in the background are mainly from the on-shell $Z$-boson. A mass-window cut on $m_{bb}$, 
$\Delta M \equiv \left| m_{bb}-m_H \right|\leq 5~{\rm GeV}$, suppresses the background dramatically. For instance, less than 1\% of the background remains after the $\Delta M$ cut. A large anomalous coupling, e.g. $\mathcal{F}_{Z\gamma}=+1$,  could lead to a few hundreds of the signal events after all the cuts and is testable experimentally. The significance ($\mathcal{S}/\sqrt{\mathcal{B}}$) increases with $\sqrt{s}$ owing both to the non-renormalizable feature of the high-dimensional operators and  also to the decreasing SM backgrounds.

Demanding the $5\sigma$ significance, $\mathcal{S}=5\sqrt{\mathcal{B}}$, yields the discovery potential of the $HZ\gamma$ and $H\gamma\gamma$ couplings in the scattering of $e^+e^- \to H\gamma$. Figure~\ref{fig:corr2} displays the discovery region of $R_\sigma$ (bold-gray band) when one anomalous coupling is considered at a time. The two-fold ambiguity of $\mathcal{F}_{Z\gamma}$ in the  measurement of $\Gamma(H\to Z\gamma)$ can be fully resolved by precise knowledge of $R_\sigma$ if $|\mathcal{F}_{Z\gamma}|$ is large enough to reach a discovery at the $e^+e^-$ collider.  The discrimination power of the two-fold $R_\sigma$ for a fixed $R_{Z\gamma/\gamma\gamma}$ increases dramatically with $\sqrt{s}$; for example, for $R_{Z\gamma}=9$, $R_\sigma$ is equal to 24 and 45 at a $\sqrt{s}=350~{\rm GeV}$ collider while it is equal to 40 and 110 at a $\sqrt{s}=1000~{\rm GeV}$ collider. We note that, the two solutions of $\mathcal{F}_{Z\gamma}$ of $R_{Z\gamma} \simeq 1$, the no-new-physics solution $\mathcal{F}_{Z\gamma}\simeq 0$ and the faked-no-new-physics solution $v^2/\Lambda^2 \mathcal{F}_{Z\gamma} \simeq -2\mathcal{F}_{Z\gamma}^{\rm SM}$, can be fully resolved in the $H\gamma$ production when $\sqrt{s}\geq 500~{\rm GeV}$. The $R_{\gamma\gamma}$ is highly limited by the current LHC data. It yields two solutions of $\mathcal{F}_{\gamma\gamma}$: one is  $v^2/\Lambda^2 \mathcal{F}_{\gamma\gamma} \sim -2 \mathcal{F}^{\rm SM}_{\gamma\gamma}$ which could be detected in the $H\gamma$ production for $\sqrt{s}\geq 500~{\rm GeV}$, the other is $\mathcal{F}_{\gamma\gamma} \sim 0 $ which cannot be probed. It turns to be slightly complicated when the $HZ\gamma$ and $H\gamma\gamma$ couplings both contribute to the $H\gamma$ production. Assuming an universal NP scale $\Lambda$, one is still able to determine the value of $\mathcal{F}_{Z\gamma}$ and $\mathcal{F}_{\gamma\gamma}$ from the measurements of $R_\sigma$ and $R_{Z\gamma/\gamma\gamma}$ and their correlations; see Eqs.~\ref{eq:xsection} and~\ref{eq:xsection2}.

Now we turn to the effective Lagrangian discussion. The new physics effects are described by an effective Lagrangian of the form~\cite{Buchmuller:1985jz, Cao:2006rn}
\beq
\mathcal{L}_{\rm eff} = \mathcal{L}_{\rm SM} + \frac{1}{\Lambda^2} \sum_i
\left(c_{i}\mathcal{O}_{i}+h.c.\right)+O\left(\frac{1}{\Lambda^{3}}
\right),
\eeq
where the coefficients $c_{i}$'s are numerical constants parametrizing the strength of the nonstandard interactions. 
The relevant dimension-6 operators for our study are $\mathcal{O}_{WW}=\left(\phi^{\dagger}\phi\right) W^I_{\mu\nu} W^{I\mu\nu}$, $\mathcal{O}_{BB}=\left(\phi^{\dagger} \phi \right) B_{\mu\nu} B^{\mu\nu}$, $\mathcal{O}_{BW}=\left(\phi^{\dagger}\tau^I\phi\right) B_{\mu\nu} W^{I\mu\nu}$ and $\mathcal{O}_{\phi\phi}=\left(D_\mu\phi\right)^\dagger\left(D^\mu\phi\right)\phi^\dagger\phi$, where $\phi$ denotes the SM scalar Higgs doublet, $W^I_{\mu\nu}$  and $B_{\mu\nu}$ are the field-strength tensors of the $SU(2)_L$ and $U(1)_Y$ gauge bosons, respectively, and $\tau^I=\sigma^I/2$ is the usual $SU(2)_L$ generator in the fundamental representation.

\begin{figure}[b]
\includegraphics[scale=0.33]{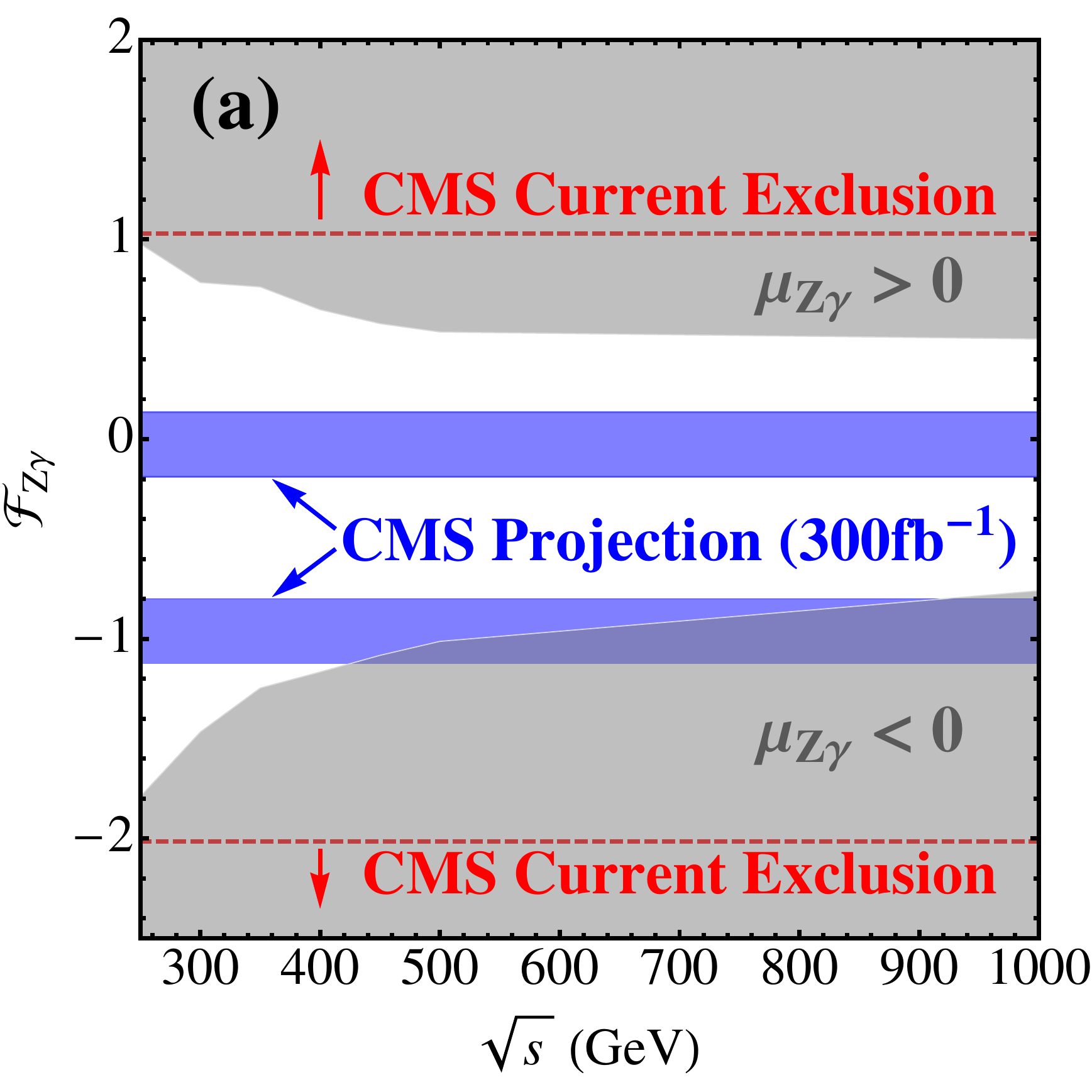}
\includegraphics[scale=0.35]{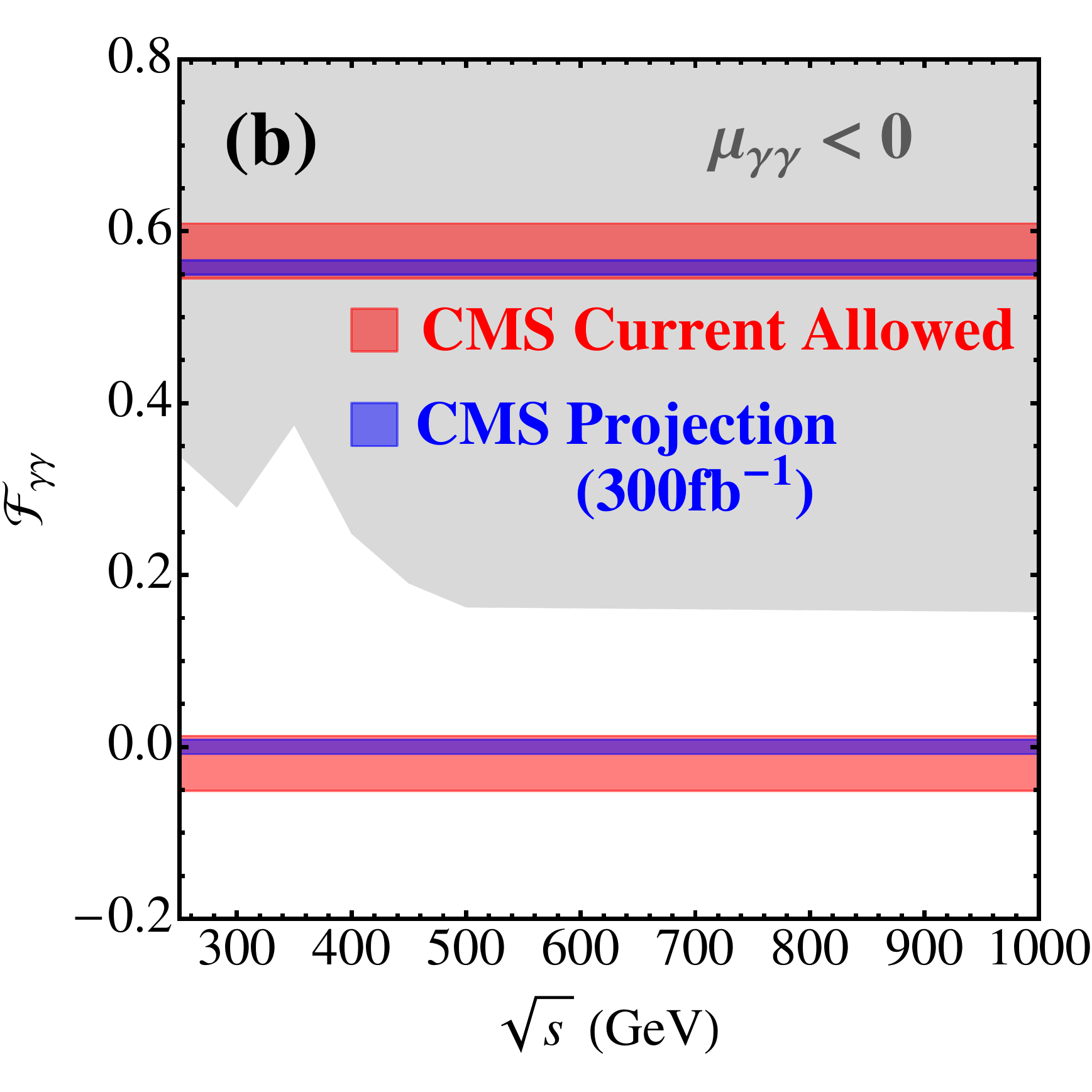}
\caption{\it Exclusion bounds (gray region) on $\mathcal{F}_{Z\gamma}$ (a) and $\mathcal{F}_{\gamma\gamma}$ (b) at the $e^+ e^-$ collider as a function of $\sqrt{s}$ for $\mathcal{L}=1~{\rm ab}^{-1}$ and $\Lambda=2~{\rm TeV}$. For comparison the CMS current limit (red) and CMS projections (blue) are also plotted.}
\label{fig:potential}
\end{figure}

The $\mathcal{O}_{\phi\phi}$ and $\mathcal{O}_{BW}$ are constrained strongly by electroweak precision measurements~\cite{Achard:2004kn,Hankele:2006ma} and are neglected in our study. After spontaneously symmetry breaking the other two operators are related to $\mathcal{F}_{Z\gamma/\gamma\gamma}$ as follows:
\bea
\mathcal{F}_{\gamma\gamma} & = &  c_{WW}\sin^2\theta_W+c_{BB}\cos^2\theta_W, \nn\\
\mathcal{F}_{Z\gamma} & = & \left( c_{WW}-c_{BB}\right)\sin(2\theta_W) .
\eea
The measurements of the $H\gamma$ production and Higgs rare decay width could probe both size and sign of $c_{WW}$ and $c_{BB}$. For example, (i) $c_{WW}\simeq -3 c_{BB}<0$ for $\mathcal{F}_{\gamma\gamma}=0$ and $\mathcal{F}_{Z\gamma}<0$; (ii) $c_{WW}=c_{BB}$ for $\mathcal{F}_{Z\gamma}=0$, $c_{WW/BB}>0$ for $\mathcal{F}_{\gamma\gamma}>0$ while $c_{WW/BB}<0$ for $\mathcal{F}_{\gamma\gamma}<0$; (iii)  $c_{WW}<c_{BB}\lesssim -c_{WW}/3$ and $c_{WW}<0$ for $\mathcal{F}_{Z\gamma}<0$ and $\mathcal{F}_{\gamma\gamma}<0$; (iv) $c_{BB}>c_{WW}>0$ for $\mathcal{F}_{Z\gamma}<0$ and $\mathcal{F}_{\gamma\gamma}>0$,  etc. Those relations would shed light on new physics searches.

When $\mathcal{F}_{Z\gamma}\sim \mathcal{F}_{\gamma\gamma}\simeq 0$, no excess of $R_\sigma$ would be observed. One can constrain tightly on $\mathcal{F}_{Z\gamma}$ and $\mathcal{F}_{\gamma\gamma}$, however. Figure~\ref{fig:potential} displays the exclusion region of $\mathcal{F}_{Z\gamma/\gamma\gamma}$ at 95\% confidence level (gray region) as a function of $\sqrt{s}$.   For comparison we also plot the CMS current limits (red) and the CMS projections (blue) at a high luminosity LHC~\cite{CMS:2013xfa}. The faked-no-new-physics scenario, $R_{Z\gamma/\gamma\gamma} \sim 1$ but $v^2/\Lambda^2 \mathcal{F}_{Z\gamma/\gamma\gamma} \sim -2 \mathcal{F}^{\rm SM}_{Z\gamma/\gamma\gamma}$, can be excluded at a high energy electron-positron collider. For example, the blue region of $\mathcal{F}_{Z\gamma}\sim -1$ in Fig.~\ref{fig:potential}(a) can be completely excluded by the $H\gamma$ production; while the red and blue regions of $\mathcal{F}_{\gamma\gamma} \sim 0.6$ in Fig.~\ref{fig:potential}(b) can also be excluded by the $H\gamma$ production.

\noindent {\it Acknowledgement:~}The work is supported in part by the National Science Foundation of China under Grand No. 11275009.

\bibliography{bibfile}
\bibliographystyle{apsrev}

\end{document}